\documentclass[final,3p,times,twocolumn]{elsarticle}
\usepackage{graphicx}
\usepackage{amssymb}
\usepackage{numcompress}
\usepackage{url}
\journal{Journal of Nanotechnology}

\begin{document}
\begin{frontmatter}

\title{Magnetic anisotropy at nanoscale}
\author{Marek W.~Gutowski}
\ead{Marek.Gutowski@ifpan.edu.pl}
\address{%
Institute of Physics, Polish  Academy  of  Sciences,  Al.~Lotnik\'ow 
32/46,  02--668 Warszawa, Poland
}%

\begin{abstract}
Nanoscale objects often behave differently than their `\/normal-sized\/' counterparts.\
Sometimes it is enough to be small in just one direction to exhibit unusual
features.\  One example of such a~phenomenon is a~very specific in-plane magnetic
anisotropy observed sometimes in very thin layers of various materials.\  Here we recall
a~peculiar form of the free energy functional nicely describing the experimental findings
but completely irrelevant and thus never observed in larger objects.
\end{abstract}



\end{frontmatter}



\section{Intriguing experimental observations}\label{breakthrough} 
In \cite{Fassbender} we find the experimentally observed in-plane magnetic anisotropy
energy (MAE) diagrams for multilayer structure
Cr$(4)$/Fe$(2)$/Cr$(d_{\mathrm{Cr}})$/Fe$(4)$/Cr$(2)$,
where numbers are thicknesses of components, expressed in~nm.\
The thickness of the middle Cr layer, $d_{\mathrm{Cr}}$, was varied in few nm range,
and the complete structures were deposited on Si$(100)$ substrate, covered with natural
SiO$_{2}$ layer $1.5$--$2.0\,$nm thick.\  The substrate was not perfectly flat --- as
a~result of ion beam erosion it was covered with quite well ordered ripples (see
atomic force microscopy (AFM) images of the substrates, Fig.~1 in
\cite{Fassbender}).\ The metallic layers were deposited using molecular beam
epitaxy (MBE) technique.\ Their transmission electron microscopy (TEM) cross
sections revealed mostly amorphous structure with small inclusions of polycrystalline
character.\ The values of MAE were derived from hysteresis loop area observed while
exciting field was oriented along successive in-plane directions.

In principle, samples of this kind should not exhibit any in-plane magnetic anisotropy.\
This is indeed the case when the substrate is flat (see Fig.~4a in~\cite{Fassbender}).\
On the rippled substrate, however,  this is no longer true and the sample exhibits
peculiar two-fold in-plane anisotropy (coercive field, Fig.~4b in~\cite{Fassbender},
MAE --- in Fig.~4c).\ It is peculiar since it is not the uniaxial anisotropy: four maxima
are visible instead of just two.

\section{The surface magnetic anisotropy of a cylinder}\label{theory} 
Consider the static configuration of individual spins located on a~surface
of a~long (ideally: infinitely long), hollow ferromagnetic cylinder.\
In absence of any external field one may expect that individual spins
may adopt one of the three stable configurations:
\begin{itemize}
\item\vskip -0.5ex
they all may be aligned with $C_{\infty}$ symmetry axis of a~cylinder.\
This is the lowest exchange energy configuration.
\item\vskip -0.5ex
they all may be oriented perpendicularly to the above symmetry axis.\
Now the exchange energy is no longer at its global minimum.\
Nevertheless, such a~configuration is stable since it realizes a local minimum
of exchange energy.
\end{itemize}
In the second case we may again distinguish two cases: either individual
spins are aligned with local $C_{2}$ symmetry axis (there are infinitely
many of them, each perpendicular to $C_{\infty}$) thus pointing inwards
or outwards of a~cylinder and being perpendicular to the cylinder's surface,
or they can be perpendicular to the local $C_{2}$ axis (laying on the cylinder's
circumference), making a~ring-shaped configuration, and producing no net
magnetization.\ 
It is easy to see that both those configurations are energetically equivalent,
since the angle between any two neighboring spins is exactly the same.

Here we concentrate only on this part of magnetostatic energy which originates
from Heisenberg-type exchange interactions between spins.\  For a pair of
nearest neighbor ($nn$) spins, say $i$ and $j$, we have
\begin{equation}
E=-2J\,\vec{S}_{i}\cdot\vec{S}_{j} = -2JS^{2}\cos\varphi_{ij},\label{nn-pair}
\end{equation}
where $J$ is an exchange integral,\ $\varphi_{ij}$ is the angle
between spins ${i}$ and ${j}$, and $S=\left|\vec{S}_{i}\right|=\left|\vec{S}_{j}\right|$.\
As the angle $\varphi_{ij}$ between neighboring spins is small, then the following
approximation is valid:
\begin{equation}
\cos\varphi_{ij} \approx 1-\varphi_{ij}^{2}/2 = 1 - \frac{1}{2}\left(\frac{\delta}{\rho}\right)^{2},\label{approx-cos}
\end{equation}
where $\delta$ is the spacing between $nn$ spins, and $\rho$ is the cylinder radius.

Full magnetostatic energy of a sample is of course the sum, running over all the pairs
of $nn$, of expressions like (\ref{nn-pair}).\ Anisotropy characterizes differences
of free energy between various directions of an external field, so any constant terms
are meaningless and may be dropped.\  In our case such a~term is ``$1$''
in (\ref{approx-cos}).\  After this is done, the exchange energy for a single $nn$ pair
of spins reads:
\begin{equation}
E\approx J\,\frac{S^{2}\delta^{2}}{\rho^{2}}\label{simple}
\end{equation}

As the sum of expressions of type (1) is hard to treat analytically,
we replace it with appropriate integral, i.e. we assume the continuous distribution
of interacting spins but we do not approximate anything else.\ Particularly, we do
not make use of approximation (\ref{approx-cos}).\ This way we have to integrate
proper expression along the elliptical path, being a~trace of a cylinder's
cross-section by a~plane parallel to the external magnetic field.\  The final result
for the surface part of the free magnetostatic energy density, $E_{\mathrm{s}}$,
already presented some time ago in \cite{moje}, reads:
\begin{equation}
E_{\mathrm{s}} = K_{\mathrm{s}}\,\left|\,\cos\theta\,\right|\label{surface}
\end{equation}
Here $K_{s}$ is the surface anisotropy constant, and $\theta$\ denotes the angle
between the direction of sample's magnetization and easy direction $C_{\infty}$.\
As one might expect\ $K_{s}\,\propto\,J/\rho^{2}$ -- in full accordance with simplified
approach, sketched in (\ref{simple}).

A~comment is in order in this place.\  Magnetocrystalline anisotropy energy density
is always expressed by even powers of $\cos\theta$  and is always a~smooth function
of the external field orientation.\  Here we have $\cos\theta$ in first power, and, additionally,
the energy density is not a~smooth function.

\section{Experimental confirmation}
The formula (\ref{surface}) has been first derived to interpret ferromagnetic resonance (FMR)
spectrum of  Co$_{68}$Mn$_{7}$Si$_{10}$B$_{15}$ glass-coated amorphous single
microwire with diameter roughly equal to $16.5\,\mu$m.\   The spectrum,
taken at fixed frequency,\ and containing more that one absorption line,\ 
could not be described (modeled) satisfactorily with conventional two- and 
fourth-order uniaxial anisotropies alone \cite{pss}.\
Unfortunately, even the inclusion of the surface anisotropy term (\ref{surface})
into the full expression for the free energy density didn't help much.\  This applies also
to further experiments, performed on similar but thinner wires, down to the diameter
of $6\,\mu$m.\
Some qualitative features of the spectra (e.g. broadening and distortion of absorption lines
at special orientations), however, could be attributed to the presence of a~non-smooth
surface anisotropy term.\  Nevertheless, it had to be concluded that the wire's diameter
was most likely to big to \emph{clearly} observe the surface anisotropy contribution.\
By the way, due to the presence of a~glassy cover, other effects, notably the
magnetostriction of the inhomogeneously stressed sample, were dominating in this
experiment.

The definitive confirmation of validity of formula (\ref{surface}) appeared only recently,
when the paper \cite{Fassbender} was published.\  Its authors admit the discrepancy
between their model of magnetic anisotropy arising at the interface between two magnetic
layers and the experimental data.\  Specifically,
they expected  \emph{a quadratic sinusoidal angle dependence} but observed
\emph{additional peaks at $\varphi=90^{\circ}$ and $\varphi=270^{\circ}$}, see Fig.~\ref{thin}
in this paper.\  Their model mimics quite well the major part of data and curves
presented in Fig.~4c \cite{Fassbender} but fails to explain the presence of those
mysterious `additional peaks.'\

\section{Discussion}
Rippled surfaces are well known in experimental practice.\ Some studies were already
performed aiming to gain the full control on ripple formation process on various substrates
\cite{Hua} (sapphire), \cite{Mollick,Sinha} (silicon), \cite{Naik} (ZnO), or to investigate
the influence of ripples on various physical properties, most notably the magnetic anisotropy,
exchange bias \cite{FassbenderR}, or morphology of magnetic domains.\ 
Recently many papers are devoted to rippled surfaces of diluted magnetic semiconductors
(DMS), with $\mathrm{(Ga,\,Mn)\,As}$\ being probably the most frequently studied
substance in this class \cite{JAP,GaMnAs}.\  The active area of research, both theoretical 
\cite{Bogdanov,Roessler} and experimental \cite{Furdyna} are competing anisotropies,
uniaxial and tetragonal, present in thin layers of this compound. 

The rippled surfaces were approximated in literature in many ways, usually as a~train
of sinusoidal waves, or as a~periodic series of Gaussian-shaped peaks or as a~periodic
set of flat islands.\  Here we propose yet another approach, namely the rippled surface
may be seen as being build by many identical, infinitely long half-cylinders, aligned
parallel to each other.\   Obviously, the period of such a~structure is equal to $4\rho$,
where $\rho$ is, as~previously, the individual cylinder's radius.\ Additionally, we neglect
eventual interactions between cylinders.

We test our theory, given in sec.~\ref{theory}.\ Using scanned data from Ref.~\cite{Fassbender},
we try to fit them to the expression
\begin{equation}
E(\theta)=a+K_{\mathrm{s}}\,|\cos\theta\,| + K_{\mathrm{u}}\cos^{2}\theta,\label{free} 
\end{equation}
i.e. taking into account only the surface anisotropy and conventional uniaxial anisotropy.\
The constant $a$ is irrelevant, but has to be fitted in order to simulate experimental data
correctly.\
It is therefore not reported in Tab.~\ref{numbers}, where the results for two available samples,
with different thicknesses $t$, are shown.\  The ripple's period was reported in \cite{Fassbender}
as being equal to $22\,$nm, so the estimated mean radius of curvature is $5.5\,$nm.\
This should be compared with the diameter of microwires used in \cite{moje}.\  Looking
at the relation (\ref{simple}), it is easy to see why the surface anisotropy term could not be
detected in earlier experiments: now the squared radius of curvature is some
$3.0\times\,\!10^5$---$2.25\times\,\!10^6$ times lower, and so many times the
expected magnitude of the effect should increase.
\begin{table}[ht]
    \caption{\label{numbers} Fitted parameters of expression (\ref{free})}
\vskip 1.0ex
\begin{tabular}{@{}rrrrrr}
\hline
 $t$   &  $\!K_{\mathrm{s}}$ &  $\!\sigma(K_{\mathrm{s}})$ & $\!K_{\mathrm{u}}$ &
 $\!\sigma(K_{\mathrm{u}})$ & $\!\left|K_{\mathrm{s}}/K_{\mathrm{u}}\right|$ \\
 nm &  $\mathrm{kJ/m}^{3}$ &  $\mathrm{kJ/m}^{3}$  &  $\mathrm{kJ/m^{3}}$ &
 $\mathrm{kJ/m^{3}}$ & \\
\hline 
   $2.0$ &   $-180$ &  $6$     & $202$  & $5$ & $0.891$ \\
   $5.2$ &   $-161$ &  $4$     & $184$  & $3$ & $0.875$ \\
\hline
\end{tabular}\label{Table}
\end{table}
The reported uncertainties for both anisotropy
constants, $K_{\mathrm{s}}$ and $K_{\mathrm{u}}$, are most likely seriously
underestimated, some $2$---$3$ times, by our quick and dirty fit.\  It is quick
and dirty because the fitting procedure has no information concerning the
uncertainties of individual measurements, and, consequently, treats all the data
as being exact.\  Even the discretization errors, being a~result of manual
scanning procedure, go unattended.

Despite of these deficiencies, the trend is clear: the magnitudes of both anisotropy
constants slightly decrease with increasing sample's thickness.
\begin{figure}[h]
    \includegraphics[keepaspectratio,angle=-90,width=\hsize]{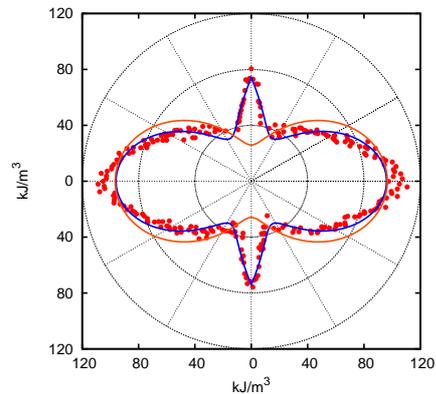}
    \caption{MAE for the sample 2.0 nm thick.\ In addition to the best fitted line
    (blue) shown is the `pure' uniaxial part (scaled) of anisotropy.\ Note the remarkable
    difference between the two near $\varphi=90^{\circ}$ and $\varphi=270^{\circ}$.\
    $156$ measurements.
    }\label{thin}
\end{figure}
The decrease of $K_{\mathrm{s}}$ probably has its roots in decreasing height
of ripples, while their
period stays unchanged during sample growth, hence the curvature radius $\rho$ effectively
increases.\  This is probably also the reason for evidently `rounded' shape of peaks visible
in Fig.~\ref{thick}.\  This feature may be also explained by finite lengths of individual
cylinders, their misalignment, or even weak, but long-range, interactions between them.\
Since it not visible so clearly in Fig.~\ref{thin}, then we should attribute it to broadening
of height (and, consequently, of curvature radii) distribution while the sample gets thicker.
\begin{figure}[ht]
    \includegraphics[keepaspectratio,angle=270,width=\hsize]{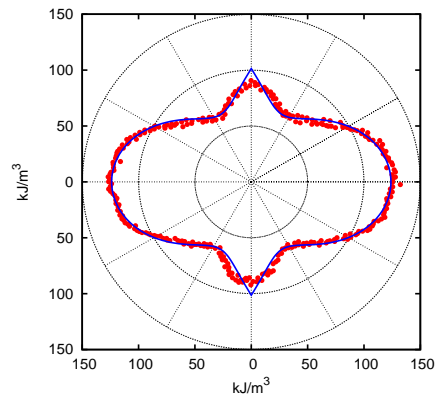}
    \caption{Same as Fig.~\ref{thin}, but for sample $5.2\,$nm thick.\ $297$ data points.\
    For clarity, the `classical' uniaxial anisotropy term is not shown.
    }\label{thick}
\label{angles}
\end{figure}
On the other hand, the drop in value of ordinary uniaxial anisotropy $K_{\mathrm{u}}$
originates most likely from the strain relaxation far from possibly mismatched substrate.

It is doubtful whether the presence of sharp, non-differentiable features, existing
in reality on any experimental curve will ever be possible to convincingly demonstrate
using the data alone.\ O${}^{\prime}$Grady \cite{OGrady} shows, using symmetry arguments,
that angular variation of many magnetic properties may be described either as even
or as odd series of $\cos\theta$.\  Similar ideas, related to variation of coercivity or
exchange bias field, were presented even earlier \cite{Ambrose}.\
Here we show one more possibility: the surface anisotropy is described by a~single
$\left|\,\cos\theta\,\right|$ term, rather hard to approximate by only few terms of
even cosine series.

It remains to be explained why $K_{\mathrm{s}}$ in Tab.~\ref{Table} is expressed
in kJ/m$^{3}$ rather than in kJ/m$^{2}$.\ This is intended, as it illustrates
convincingly (in last column) comparable shares of both types of anisotropy
in free energy (\emph{not} its density!).\
In fact, what we present there is the quantity $K_{\mathrm{s}} =
K^{\prime}_{\mathrm{s}}/t$, where  $K^{\prime}_{\mathrm{s}}$
is the true surface anisotropy, expressed in kJ/m$^{2}$, as it should, and $t$
is the sample thickness.\ Taking this into account, we have
$K^{\prime}_{\mathrm{s}}(2.0\;\mathrm{nm})=-0.36\;\mathrm{erg/cm}^{2}$ and
$K^{\prime}_{\mathrm{s}}(5.2\;\mathrm{nm})=-0.84\;\mathrm{erg/cm}^{2}$, respectively.\
One may wonder why the two estimates differ so much.\ It is less surprising
when we compare the samples' thickness ($2.0$ and $5.2\;$nm) --- in both cases
smaller than the radii ($5.5\;$nm) of our hypothetical half-cylinders.\
This means that our cylinders must be far from perfect, they are most likely
flattened, what certainly affects their curvature radii, $\rho$.\ 
Nevertheless, the values of $K_{\mathrm{s}}$ decrease with sample thickness,
as expected.\
Published values of $\left|K^{\prime}_{\mathrm{s}}\right|$ are scarce, ranging from
$0.032\;\mathrm{erg/cm}^{2}$ to as high as $1.17\;\mathrm{erg/cm}^{2}$
for Fe deposited on GaAs \cite{JMMM}.\ Our result is of the same order of magnitude.

Let us now estimate the exchange energy per single Fe--Fe pair.\
The density of elementary cells on $(001)$ surface of $\alpha$-Fe is
$n=1/a^{2}\approx1.214\times10^{19}\mathrm{m}^{-2}$, where $a=2.870\;\AA$
is $\alpha$-Fe lattice constant.\  Therefore the exchange energy, $E_{\mathrm{ex}}$,
per $a\times{a}$ square element of a~surface is $K_{\mathrm{s}}^{\prime}/n$, i.e.
$-2.965\times10^{-23}\;$J for thinner, and $-6.853\times10^{-23}\;$J
for thicker sample.\  From formula (\ref{simple}) we get
$J=E_{\mathrm{ex}}\left(\frac{\rho}{S\delta}\right)^{2}$, that is
$-3.937\times10^{-21}\;$J and $-9.099\times10^{-21}\;$J, respectively, when
$S=2.22\;$[$\mu_{\mathrm{B}}$] and $\delta=a\sqrt{3}/2\approx{0.215}\;$nm.\
Those values should still be divided by the number of $nn$ Fe pairs (4) residing
in a said $a\times{a}$ surface element.\  This is because the nearest neighbor
for any given Fe surface atom is the one laying deeper, inside the elementary
cell -- as pure iron has $bcc$ structure.\ This fact has been already accounted
for by expressing $\delta$ ($nn$ spacing) as an appropriate fraction of the
lattice constant.\ Finally we obtain $J(2.0\;\mathrm{nm})=-0.98\times10^{-21}\;$J,
and $J(5.2\;\mathrm{nm})=-2.28\times10^{-21}\;$J.\  For comparison, Ref.~\cite{PhD}
quotes $J=-1.21\;\times\;10^{-21}\;$J for pure $\alpha$-iron.\
The correspondence is amazingly good, especially that our model completely
neglects RKKY-type exchange, certainly present there, and the estimate is made
as if the surface was perfectly flat.

\section{Conclusions}
The surface anisotropy form, presented here, seems to explain the observed
features of magnetic anisotropy energy simply formidably.\ The shape of angular
dependence of MAE is reproduced much better than by any other model.\ The deduced
values of $nn$ exchange coupling strength are in good agreement with those obtained
independently.\ Moreover, they are in full accordance with intuitive understanding
what makes the surface layer: no more than two crystal planes are involved.\
Consequently, the surface layer thickness is lower than the size of a~unit cell.\
Yet, such effect can be easily observed only at nanoscale, that is in samples
thin enough.\ Only then its magnitude is comparable with ordinary uniaxial anisotropy
(see the last column of Tab.~\ref{numbers}).\  One may expect that, at least
in the case of iron, a~$\sim\!\!{500}\;$nm layer is thick enough to effectively mask
surface anisotropy effects.\

It is amazing that our original, idealized  model of non-interacting, infinitely
long half-cylinders, works so well.\ It is likely that the presence of elongated,
but finite length structures, present on nominally flat surfaces, even those
obtained by MBE technique, is sufficient to generate this form of anisotropy.\
On the other hand, it is doubtful whether it will ever be used to determine
some parameters it depends on.\ It is because the presented surface anisotropy
term is rather sensitive to the fine details of a~surface.\ Those are probably
easier to investigate using one of microscopic techniques.\  Nevertheless, using
its peculiar angular behavior, and treating it as a~`background' of known shape,
one should be able to determine important material's parameters with better accuracy
than it was possible earlier.\

The presence of non-negligible surface anisotropy, generated by surface
curvature, in addition to the edge-related effects, will affect the operation
of future spintronic devices.


\section*{Acknowledgments}
The author is very indebted to Dr. Ryszard \.Zuberek for exposing him to
problems of surface magnetism.\ Special thanks go to the unknown referee,
who's innocent remark influenced this paper very positively.\
This work was supported in part
by Polish MNiSW 2048/B/H03/2008/34 grant.

\end{document}